\documentclass[prl,pacs,twocolumn]{revtex4}

\usepackage{amsfonts}
\usepackage{amsmath}
\usepackage{amssymb}
\usepackage{bm}
\usepackage[dvips]{color}

\usepackage{graphicx}
\usepackage{subfigure}
\newcommand{\la}{\langle}
\newcommand{\ra}{\rangle}
\newcommand{\hide}[1]{}

\begin{document}


\title{Dynamics of a Magnetic Needle Magnetometer: Sensitivity to Landau--Lifshitz--Gilbert Damping}

\author{Y. B. Band$^{1,2}$, Y. Avishai$^{2,3,4}$, Alexander Shnirman$^{3,5,6}$}
\affiliation{
$^{1}$Department of Chemistry, Department of Physics, Department of Electro-Optics, and the Ilse Katz Center for Nano-Science, \\
Ben-Gurion University, Beer-Sheva 84105, Israel\\
$^{2}$New York University and the NYU-ECNU Institute of Physics at NYU Shanghai, 3663 Zhongshan Road North, Shanghai, 200062, China\\
$^{3}$Department of Physics, and the Ilse Katz Center for Nano-Science, \\
Ben-Gurion University, Beer-Sheva 84105, Israel\\
$^{4}$Yukawa Institute for Theoretical Physics, Kyoto, Japan\\
$^{5}$Institut f\"{u}r Theorie der Kondensierten Materie,
Karlsruhe Institute of Technology, D-76128 Karlsruhe, Germany\\
$^{6}$Institute of Nanotechnology, Karlsruhe Institute of Technology, D-76344
Eggenstein-Leopoldshafen, Germany
}


\begin{abstract}
An analysis of a single-domain magnetic needle (MN) in the presence of an external magnetic field ${\bf B}$ is carried out with the aim of achieving a high precision magnetometer.  We determine the uncertainty $\Delta B$ of such a device due to Gilbert dissipation and the associated internal magnetic field fluctuations that give rise to diffusion of the MN axis direction ${\bf n}$ and the needle orbital angular momentum.  The levitation of the MN in a magnetic trap and its stability are also analyzed. 
\end{abstract}


\maketitle

A rigid single-domain magnet with large total spin, e.g., $S \simeq 10^{12} \hbar$, can be used as a magnetic needle magnetometer (MNM). Recently Kimball, Sushkov and Budker \cite{KSB} predicted that the sensitivity of a precessing MNM can surpass that of present state-of-the-art magnetometers by orders of magnitude.  This prediction motivates our present study of MNM dynamics in the presence of an external magnetic field ${\bf B}$. Such analysis requires inclusion of dissipation of spin components perpendicular to the easy magnetization axis (Gilbert damping). It is due to interactions of the spin with internal degrees of freedom such as lattice vibrations (phonons), spin waves (magnons), thermal electric currents, etc.~\cite{Gilbert_04, LL_35}. Once there is dissipation, fluctuations are also present \cite{Callen_Welton}, and result in a source of uncertainty that can affect the accuracy of the magnetometer.  Here we determine the uncertainty in the measurement of the magnetic field by a MNM.  We also analyze a related problem concerning the dynamics of the needle's levitation in an {\it inhomogeneous} magnetic field, e.g., a Ioffe-Pritchard trap \cite{Gov}.

The Hamiltonian for a MN, treated as a symmetric top with body-fixed moments of inertia ${\cal I}_X = {\cal I}_Y \equiv {\cal I} \ne {\cal I}_Z$, subject to a uniform magnetic field ${\bf B}$ is,
\begin{equation}  \label{Ham_top_E_field}
  H = \underbrace{\frac{1}{2 {\cal I}} {\hat {\bf L}}^2 + (\frac{1}{2{\cal I}_Z}-\frac{1}{2{\cal I}}){\hat L_Z}^2}_{H_R}  \underbrace{-(\omega_0/\hbar)  (\hat{{\bf S}} \cdot \hat{{\bf n}})^2}_{H_A}
  \underbrace{ - \hat{{\boldsymbol \mu}} \cdot {\bf B}}_{H_B},
\end{equation}
where a hat denotes quantum operator.  In the rotational Hamiltonain $H_R$, ${\hat {\bf L}}$ is the orbital angular momentum operator  and $\hat L_{Z} = \hat{\bf L} \cdot \hat{\bf Z}$ is its component along the body-fixed symmetry axis.  ${\hat {\bf S}}$ is the needle spin angular momentum operator, and ${\hat {\bf n}}$ is the operator for ${\bf n}$ that is the unit vector in the direction of the easy magnetization axis.
The frequency appearing in the anisotropy Hamiltonian $H_A$ \cite{Brown_63} is $\omega_0 = 2 \gamma^2 K S/V$, where $K$ is the strength of the anisotropy, $V$ is the needle volume, and $\gamma = g \mu_B/\hbar$ is the gyromagnetic ratio, in which $\mu_B$ is the Bohr magnetron, and $g$ is the $g$-factor (taken to be a scalar for simplicity).  In the expression for the Zeeman Hamiltonian $H_B$, $\hat{{\boldsymbol \mu}} = g \mu_B {\hat {\bf S}}$ is the magnetic moment operator.  
The Heisenberg equations of motion are
\begin{eqnarray}
&&   \dot{\hat{{\bf S}}} = -g \mu_B {\bf B} \times {\hat {\bf S}} + 2\frac{\omega_0}{\hbar} ({\hat {\bf S}} \times {\hat {\bf n}}) ({\hat {\bf S}} \cdot {\hat {\bf n}}) ,  \label{dot_S_simple_4}\\
&&   \dot{\hat{{\bf L}}}\mbox{=-} 2 \frac{\omega_0}{\hbar} (\hat {\bf S} \times \hat {\bf n}) ({\bf S} \cdot \hat {\bf n}), \label{dot_L_simple_4}\\
&&   \dot{\hat{{\bf J}}} =  -g \mu_B  {\bf B}  \times \hat{{\bf S}},  \label{dot_J_simple_4} \\
&& \dot{\hat{{\bf n}}} = \frac{{\boldsymbol{\mathcal I}}^{-1}}{\hbar} 
 [{\hat {\bf L}} \times \hat{{\bf n}} \mbox{+} i \hbar  \hat{{\bf n}}],
 \label{dot_n_simple_4}
\end{eqnarray}
where $\hat{{\bf J}} = \hat{{\bf L}}+\hat{{\bf S}}$ is the total angular momentum operator and ${\boldsymbol{\mathcal I}}$ is the moment of inertia tensor.

The dynamics of a MN can be treated semiclassically because $S$ is very large.  A mean--field approximation \cite{Zobay_00, Band_07, Band_13} is obtained by taking quantum expectation values of the operator equations and assuming that for a given operator $\hat{A}$, the inequality $\sqrt{\la \hat{A}^2 \ra - \la \hat{A} \ra^2} \ll \vert \la \hat{A} \ra \vert$ holds, (an assumption warranted for large $S$).  Hence, the expectation values of a product of operators on the RHS of Eqs.~(\ref{dot_S_simple_4})-(\ref{dot_n_simple_4}) can be replaced by a product of expectation values. The semiclassical equations are equivalent to those obtained in a classical Lagrangian formulation.  Dissipation is accounted for by adding the Gilbert term \cite{Gilbert_04, Brown_63} $- \alpha {\bf S} \times ({\dot {\bf S}}/\hbar - {\boldsymbol \Omega} \times {\bf S}/\hbar)$ to the RHS of the expectation value of Eq.~(\ref{dot_S_simple_4}) and subtracting it from the RHS of Eq.~(\ref{dot_L_simple_4}). Here $\alpha$ is the dimensionless friction parameter, and the term ${\boldsymbol \Omega} \times {\bf S}$ transforms from body fixed to space fixed frames.  Note that Gilbert damping is due to {\em internal} forces, hence ${\bf J}$ is not affected and Eq.~(\ref{dot_J_simple_4}) remains intact.

It is useful to recast the semiclassical dynamical equations of motion in reduced units by defining dimensionless vectors: the 
unit spin ${\bf m} \equiv {\bf S}/S$, the orbital angular momentum  ${\boldsymbol \ell} \equiv {\boldsymbol L}/S$, the 
total angular momentum, ${\bf j} = {\bf m} + {\boldsymbol \ell}$ and the unit vector in the direction of the magnetic field 
${\bf b}={\bf B}/B$:
\begin{equation} \label{dot_S_simple_5}
    {\dot {\bf m}} = \omega_B {\bf m} \times {\bf b} + \omega_0 ({\bf m} \times {\bf n}) ({\bf m} \cdot {\bf n}) 
    - \alpha {\bf m} \times ({\dot {\bf m}} - {\boldsymbol \Omega} \times {\bf m}),
\end{equation}
\vspace{-0.3in}
\begin{eqnarray} \label{dot_L_simple_5}
    {\dot {\boldsymbol \ell}} &=& - \omega_0 ({\bf m} \times {\bf n}) ({\bf m} \cdot {\bf n}) + \alpha {\bf m} \times ({\dot {\bf m}} - {\boldsymbol \Omega} \times {\bf m}) ,
\end{eqnarray}
\vspace{-0.35in}
\begin{eqnarray} \label{dot_n_simple_5}
    {\dot {\bf n}} &=& {\boldsymbol \Omega} \times {\bf n} ,
\end{eqnarray}
\vspace{-0.35in}
\begin{eqnarray} \label{dot_j_simple_5}
    {\dot {\bf j}} &=& \omega_B {\bf m} \times {\bf b} ,
\end{eqnarray}
where the angular velocity vector ${\boldsymbol \Omega}$ is given by
\begin{eqnarray} \label{Eq:Omega}
{\boldsymbol \Omega} &=& (\omega_3 - \omega_1) ({\boldsymbol \ell} \cdot {\bf n}) {\bf n}  + \omega_1 {\boldsymbol \ell} \nonumber \\
&=& (\omega_3 - \omega_1) \left[ ({\bf j} - {\bf m}) \cdot {\bf n} \right] {\bf n}  + \omega_1 ({\bf j} - {\bf m}) .
\end{eqnarray}
Here $\omega_B = \gamma |{\bf B}|$ is the Larmor frequency, $\omega_1 = S/{\cal I}_X$, and $\omega_3 = S/{\cal I}_Z$.  Similar equations were obtained in Ref.~\cite{Keshtgar}, albeit assuming that the deviations of ${\bf n}(t)$ and ${\bf m}(t)$ from ${\bf b}$ are small.  We show below that the dynamics can be more complicated than simply precession of the needle about the magnetic field, particularly at high magnetic fields where nutation can be significant. 

\begin{figure} [ht]
\centering
\includegraphics[width=0.8\linewidth]{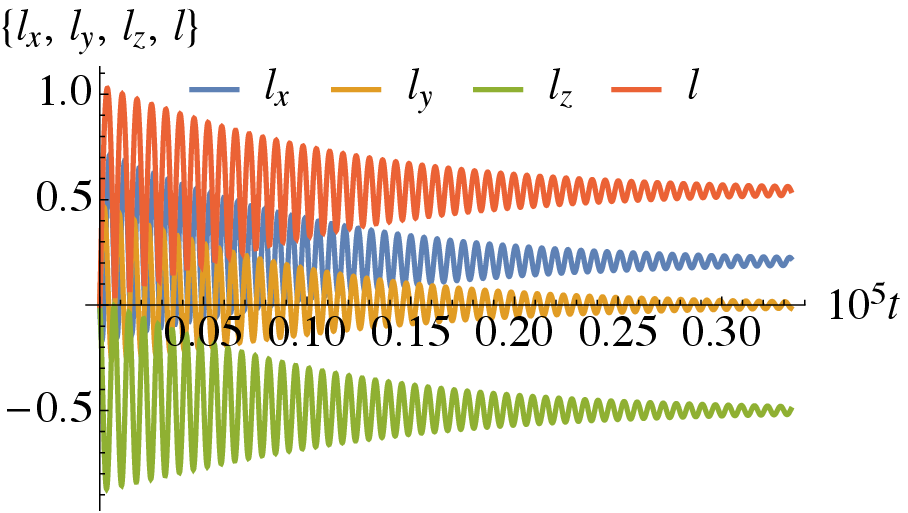}
\centering
\includegraphics[width=0.8\linewidth]{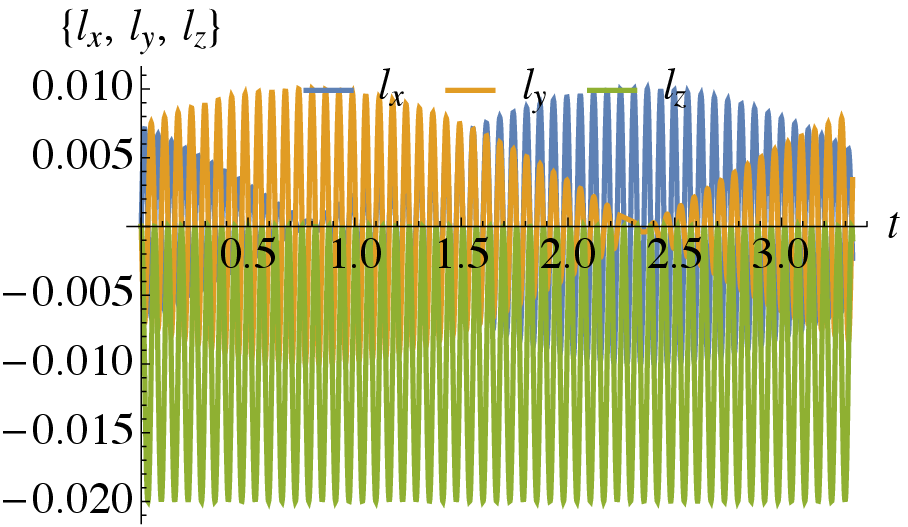}
\centering
\includegraphics[width=0.8\linewidth]{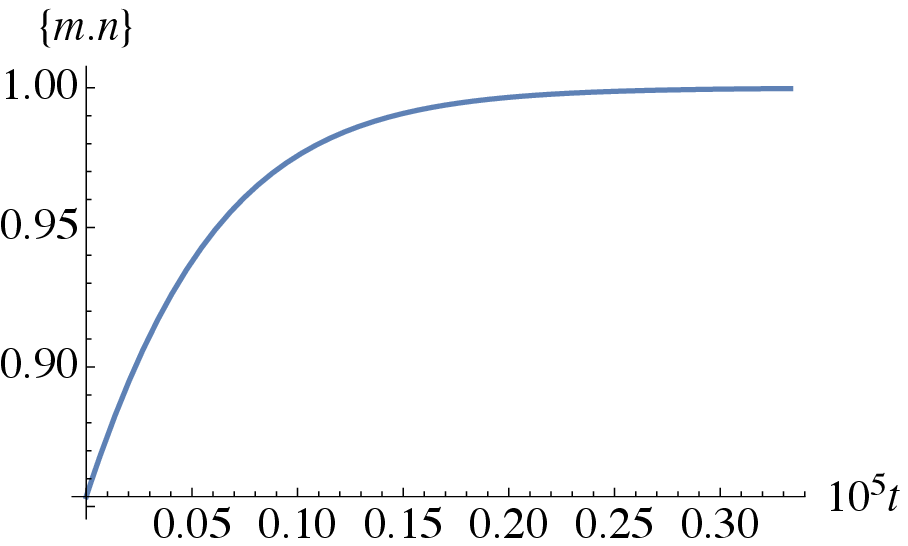}
\caption{(color online) (a) The normalized spin vector ${\bf m}$ versus time for the low-field case at short times (5 orders of magnitude shorter than in Fig.~2) when the initial spin is not along the fast axis. (b) The reduced orbital angular momentum vector ${\boldsymbol \ell}(t)$. (c) The inner product ${\bf m}(t) \cdot {\bf n}(t)$ (the projection of the spin on the fast magnetic axis of the needle.}
\label{Fig_1}
\end{figure}

For the numerical solutions presented below we are guided by Ref.~\onlinecite{KSB}, which uses parameters for bulk cobalt, and take $\omega_1 = 100$ s$^{-1}$, $\omega_3 = 7000$ s$^{-1}$, anisotropy frequency $\omega_0 = 10^8$ s$^{-1}$, Gilbert constant $\alpha = 0.01$, temperature $T=300$ K, and $N=S/\hbar=10^{12}$.
First, we  elucidate the effects of Gilbert dissipation, and consider the short time behavior in a weak magnetic field, $\omega_B = 1$ s$^{-1}$.  The initial spin direction is intentionally chosen {\em not} to be along the easy magnetic axis; ${\bf n}(0) = (1/2, 1/\sqrt{2}, 1/2)$, ${\bf m}(0) = (1/\sqrt{2}, 1/\sqrt{2}, 0)$, ${\boldsymbol \ell}(0) = (0, 0, 0)$.  Figure \ref{Fig_1}(a) shows the fast spin dissipation as it aligns with the easy axis of the needle, i.e., ${\bf m}(t) \rightarrow {\bf n}(t)$ after a short time, and Fig.~\ref{Fig_1}(b) shows relaxation of the oscillations in ${\boldsymbol \ell}(t)$, while $\ell_x(t)$ and $\ell_y(t)$ approach finite values.  Figure \ref{Fig_1}(c) shows the inner product ${\bf m} \! \cdot \! {\bf n}$, which clearly tends to unity on the timescale of the figure.  Increasing $\alpha$ leads to faster dissipation of ${\bf m}(t)$, but the short-time saturation values of both ${\bf m}(t)$ and ${\boldsymbol \ell}(t)$ are almost independent of $\alpha$.
 
\begin{figure} [ht]
\centering
\includegraphics[width=0.8\linewidth]{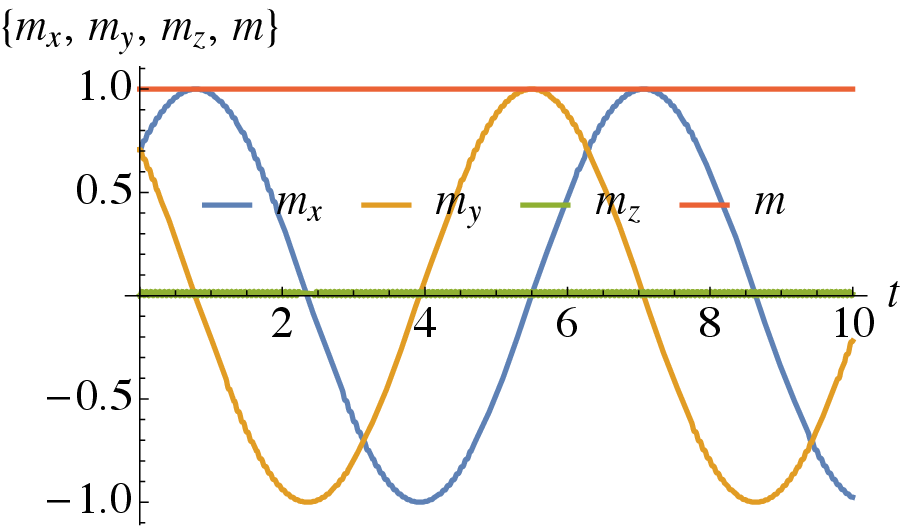}
\centering
\includegraphics[width=0.8\linewidth]{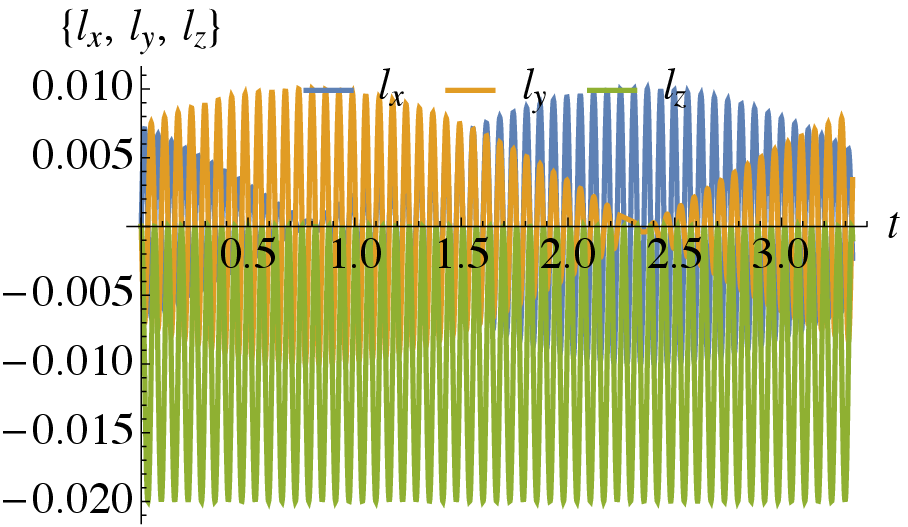}
\centering
\includegraphics[width=0.8\linewidth]{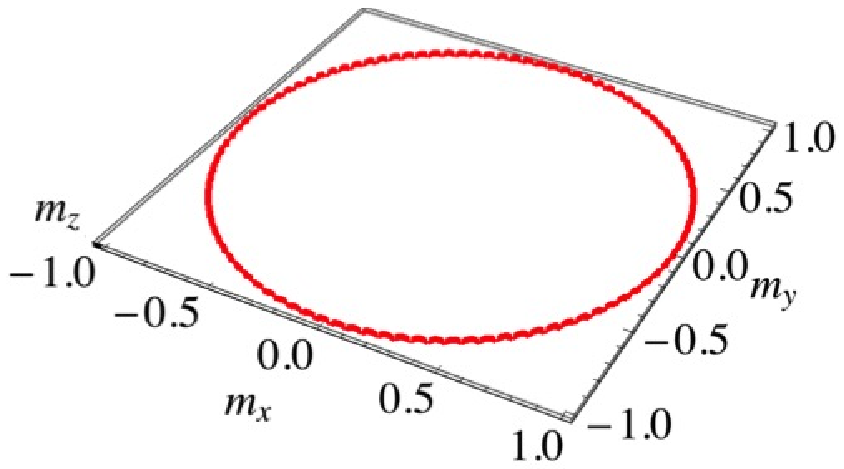}
\caption{(color online) Dynamics for the low-field case ($\omega_B = 1$ s$^{-1}$), over relatively long timescales relative to those in Fig.~\ref{Fig_1}. (a) ${\bf m}$ versus time in units of seconds (note that ${\bf n}$ is indistinguishable from ${\bf m}$ on the scale of the figure). (b) ${\boldsymbol \ell}(t)$ (note that it stays small compared to $S$).  (c) Parametric plot of the needle spin vector ${\bf m}(t)$ showing that nutation is almost imperceptible for small fields [contrast this with the large field result in Fig.~\ref{Fig_3}(c)]; only precession is important.}
\label{Fig_2}
\end{figure}

We consider now the long time dynamics (still in the weak field regime) and take the initial value of the spin to coincide with the easy magnetization axis, e.g., ${\bf m}(0) = {\bf n}(0) = (1/\sqrt{2}, 1/\sqrt{2}, 0)$, with all other parameters unchanged.  The spin versus time is plotted in Fig.~\ref{Fig_2}(a).  The unit vectors ${\bf m}(t)$ and ${\bf n}(t)$ are almost identical, and since their $z$-component is nearly zero, they move together in the $x$-$y$ plane. In this weak field case, the nutation is small, and the fast small-oscillations due to nutation are barely visible.  The orbital angular momentum dynamics is plotted in Fig.~\ref{Fig_2}(b) [note the different timescale in (a) and (b)] and shows that ${\boldsymbol \ell}(t)$ oscillates with a frequency equal to that of the fast tiny-oscillation of ${\bf m}(t)$ [the oscillation amplitude is $0.02 \, \vert {\bf m}(t) \vert $].  Figure~\ref{Fig_2}(c) shows a parametric plot of ${\bf m}(t)$ versus time.  The nutation is clearly very small; the dynamics of ${\bf m}(t)$ consists almost entirely of precession at frequency $\omega_B$.

\begin{figure} [ht]
\centering
\includegraphics[width=0.8\linewidth]{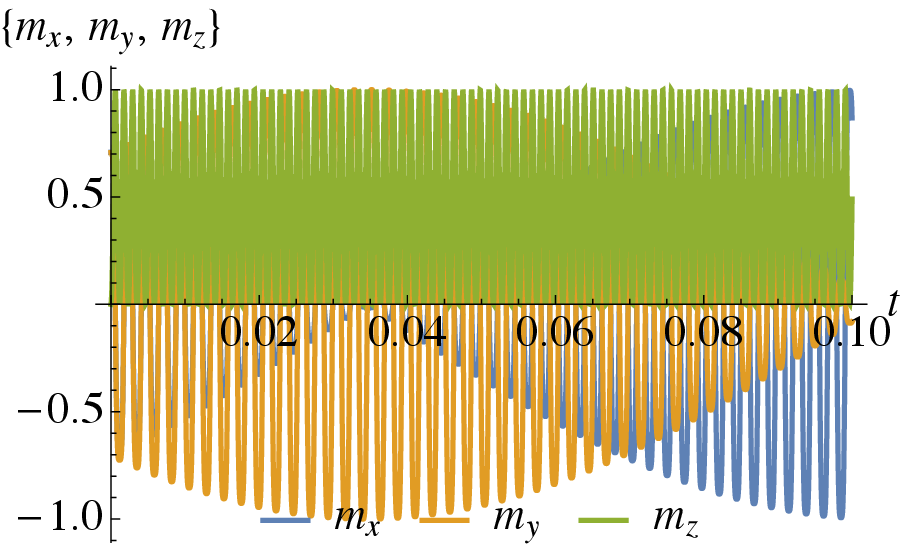}
\centering
\includegraphics[width=0.8\linewidth]{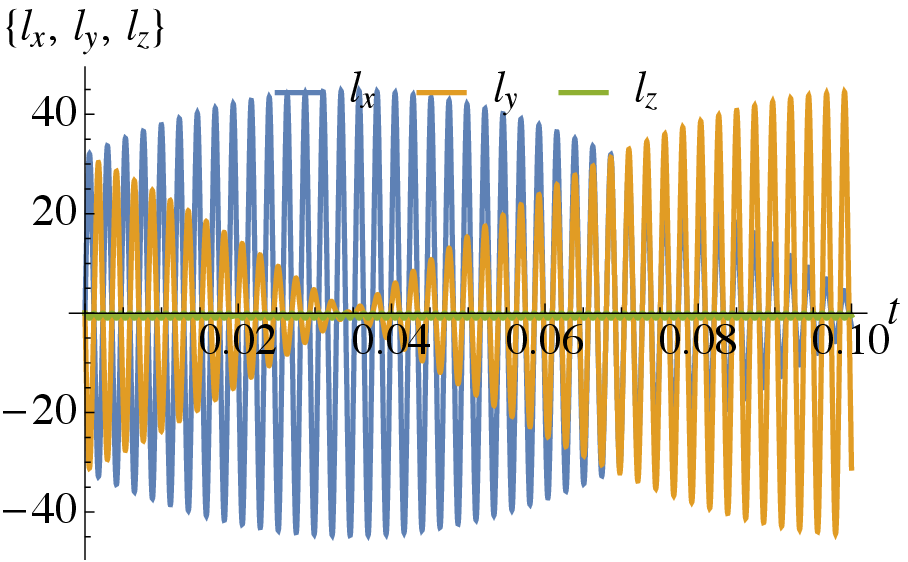}
\centering
\includegraphics[width=0.8\linewidth]{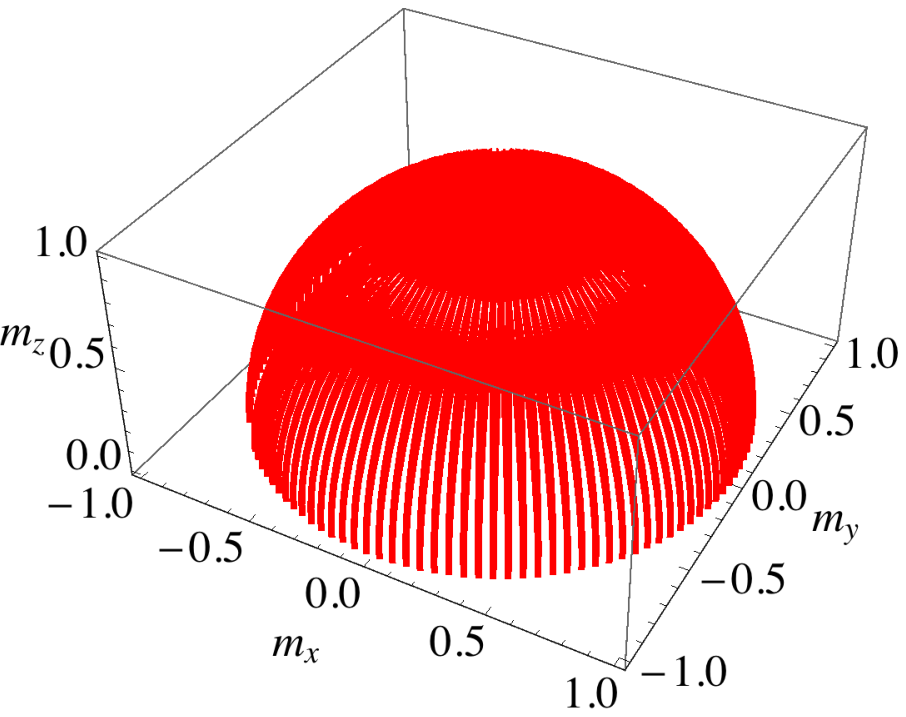}
\caption{(color online) High-field case ($\omega_B = 10^{5}$ s$^{-1}$). (a) ${\bf m}(t)$ [which is almost numerically equal to ${\bf n}(t)$]. (b) ${\boldsymbol \ell}(t)$ (note the ordinate axis scale is $[-40,40]$). (c) Parametric plot of the needle spin vector ${\bf m}(t)$ showing that strong nutation occurs for large fields in addition to precession.}
\label{Fig_3}
\end{figure}

Figure \ref{Fig_3} shows the  dynamics at high magnetic field ($\omega_B = 10^{5}$ s$^{-1}$) with all the other parameters unchanged.  Figure~\ref{Fig_3}(a) shows ${\bf m}$ versus time, and now the nutation is clearly significant.  For the high magnetic field case, ${\bf m}(t)$ is also almost numerically equal to ${\bf n}(t)$. ${\boldsymbol \ell}(t)$ is plotted in Fig.~\ref{Fig_3}(b). Its amplitude is very large, $\ell(t) \approx 40 \, m(t)$. However, its oscillation frequency is comparable with that of ${\bf m}(t)$.   In contrast with the results in Fig.~\ref{Fig_2}, here, in addition to precession of the needle, significant nutation is present, as shown clearly in the parametric plot of the needle spin vector ${\bf m}(t)$ in Fig.~\ref{Fig_3}(c).

We now determine the uncertainty of the MNM due to internal magnetic field fluctuations related to the Gilbert damping.  A stochastic force ${\boldsymbol \xi}(t)$, whose strength is determined by the fluctuation--dissipation theorem \cite{Callen_Welton}, is added to Eq.~(\ref{dot_S_simple_5}), in direct analogy with the treatment of Brownian motion where both dissipation and a stochastic force are included \cite{stochastic}:
\begin{eqnarray} \label{Eq:mdot}
{\dot {\bf m}}  &=& {\bf m} \times (\omega_B {\bf b} + {\boldsymbol \xi}) + \omega_0 ({\bf m} \times {\bf n}) ({\bf m} \cdot {\bf n}) \nonumber \\
 && -\alpha \,  {\bf m} \times ({\dot {\bf m}} - {\boldsymbol \Omega} \times {\bf m}) .
\end{eqnarray}
${\boldsymbol \xi}(t)$ is internal to the needle and therefore it does not affect the total angular momentum ${\bf j}$ directly, i.e., ${\boldsymbol \xi}(t)$ does not appear in Eq.~(\ref{dot_j_simple_5}) [since the term $-{\bf m} \times {\boldsymbol \xi}$ is also added to the RHS of (\ref{dot_L_simple_5})].  However, as shown below, ${\boldsymbol \xi}(t)$ affects $\boldsymbol \ell$ as well as $\bf m$, causing them to wobble stochastically.  This, in turn, makes $\bf j$ stochastic as well via the Zeeman torque [see Eq.~(\ref{dot_j_simple_5})].

The fluctuation-dissipation theorem \cite{Callen_Welton} implies
\begin{eqnarray}  \label{Eq:NoiseFDT}
&& \langle \xi_\alpha \xi_\beta \rangle_\omega \equiv \int dt \langle \xi_\alpha(t)\xi_\beta(0)\rangle\, e^{i\omega t} \nonumber \\
&&= \delta_{\alpha\beta}\,
\frac{\alpha \omega \coth(\hbar\omega/2 k_B T)}{N} \approx  \delta_{\alpha\beta}\, \frac{2\alpha k_B T}{\hbar N}\ ,
\end{eqnarray}
where $N = S/\hbar$, and the last approximation is obtained under the assumption that $\hbar \omega \ll k_B T$. Note that Eq.~(\ref{Eq:mdot}) should be solved together with Eqs.~(\ref{dot_n_simple_5}) and (\ref{dot_j_simple_5}).

The presence of the anisotropy term in Eq.~(\ref{Eq:mdot}) makes numerical solution difficult for large $\omega_0$.  Hence, we consider a perturbative expansion in powers of $\lambda \equiv \omega_1/\omega_0$: 
${\bf m}(t) = {\bf n}_0(t) + \lambda \, \delta{\bf m}(t) + \ldots$, ${\bf n}(t) = {\bf n}_0(t) + \lambda \, \delta{\bf n}(t) + \ldots$, ${\bf j}(t) = {\bf j}_0(t) + \lambda \, \delta{\bf j}(t) + \ldots$.
Since $\omega_0$ is the largest frequency in the problem, the inequalities $\alpha\omega_0\gg \omega_B,\omega_1,\omega_3$ hold. Moreover, the Gilbert constant $\alpha$ is large enough to effectively pin ${\bf m}(t)$ to ${\bf n}(t)$ [hence ${\bf j}(t) = {\boldsymbol \ell}(t) + {\bf m}(t) \approx {\boldsymbol \ell}(t) + {\bf n}(t)$]. Therefore, an adiabatic approximation to the set of dynamical stochastic equations can be obtained.  The zero order term in $\lambda$ reads:
\begin{equation} \label{Eq:jndot0}
    {\dot {\bf j}}_0 = \omega_B {\bf n}_0 \times {\bf b} , \quad
    {\dot {\bf n}}_0 = \omega_1 {\bf j}_0 \times {\bf n}_0 ,
\end{equation}
where ${\boldsymbol \Omega}$ was approximated by ${\boldsymbol \Omega}_0 = (\omega_3 - \omega_1) ({\bf j}_0 \cdot {\bf n}_0 - 1) {\bf n}_0  + \omega_1 ({\bf j}_0 - {\bf n}_0)$ in Eqs.~(\ref{dot_n_simple_5}) and (\ref{Eq:Omega}) in obtaining (\ref{Eq:jndot0}) \cite{gravitation}. The solution to Eqs.~(\ref{Eq:jndot0}) [for times beyond which Gilbert dissipation is significant so ${\bf m}(t) \approx {\bf n}(t)$] is very close to that obtained from Eqs.~(\ref{dot_S_simple_5})-(\ref{dot_n_simple_5}).

Expanding Eq.~(\ref{Eq:mdot}) in powers of $\lambda$ and keeping only the first order terms (the zeroth order term on the LHS vanishes since ${\bf m}_0 = {\bf n}_0$), we get:
$\omega_1 \,(\delta {\bf m} - \delta {\bf n}) \times {\bf n}_0 =  \dot{\bf n}_0 - \omega_B {\bf n}_0 \times {\bf b} +\alpha {\bf n}_0 \times ({\dot {\bf n}_0} - {\boldsymbol \Omega_0} \times {\bf n}_0) - {\bf n}_0 \times  {\boldsymbol \xi}$.
Taking Eq.~(\ref{Eq:jndot0}) into account and introducing the notation $\delta{\boldsymbol \eta} \equiv \delta {\bf m} - \delta {\bf n}$, we obtain
\begin{equation}  \label{Eq:delta_m}
\delta{\boldsymbol \eta} \times {\bf n}_0 =  {\bf j}_0 \times {\bf n}_0  - (\omega_B/\omega_1) {\bf n}_0 \times {\bf b}  -(1/\omega_1) {\bf n}_0 \times  {\boldsymbol \xi}\ ,
\end{equation}
and from Eqs.~(\ref{dot_n_simple_5}) and (\ref{dot_j_simple_5}) we find
\begin{equation} \label{Eq:deltajdot}
\frac{d}{dt}{\delta {\bf j}} =\omega_B (\delta{\bf n} + \delta{\boldsymbol \eta}) \times {\bf b}  \ ,
\end{equation}
\begin{eqnarray} \label{Eq:deltandot}
\frac{d}{dt} \delta {\bf n} &=&\omega_1 ({\bf j}_0-{\bf n}_0) \times \delta {\bf n} +\omega_1 (\delta {\bf j} -\delta{\bf n} - \delta{\boldsymbol \eta}) \times {\bf n}_0 \nonumber \\
&=& \omega_1 {\bf j}_0 \times \delta {\bf n} +\omega_1 (\delta {\bf j} -\delta{\boldsymbol \eta}) \times {\bf n}_0  \ .
\end{eqnarray}
To first order in $\lambda$, $\delta {\bf n} \perp {\bf n}_0$ (since ${\bf n}$ must be a unit vector), and $\delta {\bf m} \perp {\bf n}_0$, hence $\delta{\boldsymbol \eta} \perp {\bf n}_0$.  Therefore, $\delta{\boldsymbol \eta} \times {\bf b} =  [{\bf j}_0 -({\bf j}_0 \cdot {\bf n}_0) {\bf n}_0]  \times {\bf b} + (\omega_B/\omega_1) [{\bf b} -({\bf b} \cdot {\bf n}_0) {\bf n}_0]  \times {\bf b} + \omega_1^{-1}  [{\boldsymbol \xi} - ({\boldsymbol \xi} \cdot {\bf n}_0) {\bf n}_0]  \times {\bf b}$ on the RHS of Eq.~(\ref{Eq:deltajdot}) and
\begin{eqnarray} \label{Eq:deltajdot_new}
&& \frac{d}{dt}{\delta {\bf j}}  =  \omega_B \delta{\bf n} \times {\bf b}  + \omega_B [{\bf j}_0 -({\bf j}_0 \cdot {\bf n}_0) {\bf n}_0]  \times {\bf b} \nonumber \\
&& - \frac{\omega_B^2}{\omega_1} ({\bf b} \cdot {\bf n}_0) {\bf n}_0  \times {\bf b} + \frac{\omega_B}{\omega_1}  [{\boldsymbol \xi} - ({\boldsymbol \xi} \cdot {\bf n}_0) {\bf n}_0]  \times {\bf b} .
\end{eqnarray}
Equations~(\ref{Eq:jndot0}), (\ref{Eq:deltandot}) and (\ref{Eq:deltajdot_new}) form a closed system of stochastic differential equations [upon using Eq.~(\ref{Eq:delta_m}) to substitute for $\delta{\boldsymbol \eta} \times {\bf n}_0$ on the RHS of Eq.~(\ref{Eq:deltandot})].  With the largest frequency $\omega_0$ eliminated, a stable numerical solution is obtained. Moreover, for small magnetic field (where $\omega_B$ is the smallest frequency in the system), an analytic solution of these equations is achievable.  To obtain an analytic solution to Eqs.~(\ref{Eq:jndot0}), let us transform to the frame rotating around ${\bf B}$ with frequency $\omega_B$ 
to get equations of the form $\frac{d}{d\tau} {\bf v} = \frac{d}{d t} {\bf v}  + \omega_B {\bf b} \times {\bf v}$ (which defines $\tau$):
\begin{eqnarray} 
&&    \frac{d}{d\tau} {\bf n}_0 = - \omega_1 {\bf n}_0 \times \left({\bf n}_0  - {\bf j}_0 + \frac{\omega_B}{\omega_1} {\bf b}\right) ,
\label{Eq:ndot0_n_n} \\
&&      \frac{d}{d\tau} {\bf j}_0 = \omega_B {\bf b} \times \left({\bf n}_0  - {\bf j}_0 + \frac{\omega_B}{\omega_1} {\bf b}\right).
 \label{Eq:jdot0_n}
\end{eqnarray}
If the initial condition is ${\bf n}_0(0) - {\bf j}_0(0) + (\omega_B/\omega_1) {\bf b} = 0$, then, in the rotating frame ${\bf j}_0(\tau)$ and ${\bf n}_0(\tau)$ are constant vectors.  Note that this initial condition is only slightly different from the ``ordinary'' initial condition ${\bf n}_0(0)  = {\bf j}_0(0)$ since $(\omega_B/\omega_1)  \ll 1$ for small magnetic fields.  Hence, in the rotating frame,
\begin{eqnarray} \label{Eq:dndtau}
    \frac{d}{d\tau} \delta {\bf n} &=&  \omega_1 {\bf n}_0 \times (\delta {\bf n}  - \delta{\bf j} + \delta{\boldsymbol \eta}) ,
\end{eqnarray}
\begin{eqnarray} \label{Eq:djdtau}
     \frac{d}{d\tau} \delta {\bf j} &=& -\omega_B {\bf b} \times (\delta {\bf n}  - \delta{\bf j} + \delta{\boldsymbol \eta}) .
\end{eqnarray}
With the special initial condition being satisfied, Eq.~(\ref{Eq:delta_m}) becomes $\delta{\boldsymbol \eta} \times {\bf n}_0 =  -(1/\omega_1) {\bf n}_0 \times  {\boldsymbol \xi}$, and Eqs.~(\ref{Eq:dndtau})-(\ref{Eq:djdtau}) become a set of first order differential equations with time-independent coefficients.  Their solution for initial conditions, $\delta {\bf n}(t=0)=0$, $\delta {\bf j}(t=0)=0$ is, 
\begin{eqnarray}  \label{eq:expCsolution}
\left( \! \begin{array}{c} \delta {\bf n}(t) \\ \delta {\bf j}(t) \end{array}  \! \right) = \int\limits_0^t dt_1 
\exp\left[C(t-t_1)\right] \,C
\, \left(  \! \begin{array}{c} \delta {\boldsymbol \eta}(t_1) \\ 0 \end{array}  \! \right) \ ,
\end{eqnarray}
where the constant matrix $C =\left( \begin{array}{cc} \phantom{-}A & -A\\ -B &\phantom{-}B \end{array}\right)$ has dimension $6$$\times$$6$ and the $3$$\times$$3$ matrices $A$ and $B$ are given by $A_{ij} = -\omega_1\epsilon_{ijk} n_0^k$, $B_{ij} = -\omega_B\epsilon_{ijk} b^k$.  Without loss of generality we can choose ${\bf n}_0 = {\hat {\bf z}}$ and ${\bf b} =\omega_B( \cos\theta\, {\hat {\bf z}} + \sin\theta\, {\hat {\bf x}})$, 
where $\theta$ is the angle between the easy magnetization axis and the magnetic field.  In this basis,
$\langle \delta\eta_x  \delta\eta_x\rangle_\omega =
\langle \delta\eta_y  \delta\eta_y \rangle_\omega  \approx \omega_0^{-2} \langle \xi_x \xi_x\rangle_\omega= \omega_0^{-2} \langle \xi_y \xi_y\rangle_\omega =S_a(\omega)$, and
$\langle \delta\eta_z  \delta\eta_z \rangle_\omega = 0$. Here $\langle x x\rangle_\omega \equiv
\int dt \, e^{i\omega t} \langle x(t)x(0)\rangle$ and [see Eq.~(\ref{Eq:NoiseFDT})]
$S_a(\omega)=\frac{\alpha \omega \coth(\hbar\omega/2 k_B T)}{\omega_0^2 N} \approx \frac{2\alpha k_B T}{N \hbar \omega_0^2}$.

We are particularly interested in the quantities $\langle \delta n_y^2(t) \rangle\equiv \langle \delta n_y(t) \delta n_y(t)  \rangle$ and $\langle \delta j_y^2(t) \rangle \equiv \langle \delta j_y(t) \delta j_y(t)\rangle$ because, in the basis chosen above, the $y$-axis is the direction of precession of ${\bf n}_0$ around ${\bf b}$.  Using Eq.~(\ref{eq:expCsolution}) we obtain
$\langle \delta n_y^2(t) \rangle \approx  t\, \omega_1^2 S_a(\omega \sim \omega_1)$.
Assuming the precession of ${\bf n}$ is measured, [or equivalently, the precession of ${\bf m}$, since they differ only for short timescales of order $(\alpha\omega_0)^{-1}$], the uncertainty in the precession angle is $\langle (\Delta\varphi)^2\rangle \approx t\, \omega_1^2 S_a(\omega \sim \omega_1)$.  We thus arrive at our central result: the precision with which the precession frequency can be measured is,
$\Delta \omega_B = \frac{\sqrt{\langle (\Delta\varphi )^2\rangle}}{t} 
\approx \frac{\omega_1}{\omega_0} \sqrt{\frac{2\alpha k_B T}{\hbar N}}\,\frac{1}{\sqrt{t}}$.
Equivalently, the magnetic field precision is,
\begin{equation}
\Delta B = \frac{\Delta \omega_B}{\gamma}  \approx 
\frac{\hbar}{g\mu_B} \frac{\omega_1}{\omega_0} \sqrt{\frac{2\alpha k_B T}{\hbar N}}\,\frac{1}{\sqrt{t}}\ .
\end{equation}
For the parameters used in this paper we find $\Delta B \approx \frac{5 \times 10^{-18}}{\sqrt{t[s]}}$ Tesla (independent of $\omega_B$). This result should be compared with the scaling $\Delta B \propto t^{-3/2}$ obtained in Ref.~\onlinecite{KSB}.  Therein, the initial uncertainty of the spin direction relative to the needle axis was estimated from the fluctuation-dissipation relation and the deterministic precession resulted in the $t^{-3/2}$ scaling of the precession angle uncertainty (in addition this angle was assumed to be small). In contrast, we consider the uncertainty acquired due to Gilbert dissipation {\em during} the precession, allowing the precession angle to be large. Thus, the standard $1/\sqrt{t}$ diffusion scaling is obtained and dominates for times that are even much longer than those considered in Ref.~\onlinecite{KSB}.

In the Supplemental Material \cite{suppl} we discuss three relevant related issues. (a) The time at which diffusion stops because equipartition is reached (we estimate the time when the energy stored in stochastic orbital motion becomes of order $k_B T$). (b) The uncertainty of the magnetic field for experiments in which the fast precession of ${\bf n}$ around ${\bf j}$ is averaged out in the measurement, and the diffusion of ${\bf j}$ determines $\Delta B$. (c) We consider the related problem of the dynamics and stability of a rotating MN in an inhomogeneous field (e.g., levitron dynamics in a Ioffe-Pritchard trap \cite{Berry, levitron_movie}).

In conclusion, we show that $\Delta B$ due to Gilbert damping is very small; external noise sources, as discussed in Ref.~\cite{KSB}, will dominate over the Gilbert noise for weak magnetic fields.  A closed system of stochastic differential equations, (\ref{Eq:jndot0}), (\ref{Eq:deltandot}) and (\ref{Eq:deltajdot_new}), can be used to model the dynamics and estimate $\Delta B$ for large magnetic fields.  A rotating MN in a magnetic trap can experience levitation, although the motion does not converge to a fixed point or a limit cycle; an adiabatic--invariant stability analysis confirms stability \cite{suppl}.

\begin{acknowledgments}
This work was supported in part by grants from the DFG through the DIP program (FO703/2-1).  Useful discussions with Professor Dmitry Budker are gratefully acknowledged.  A.~S.~was supported by DFG Research
Grant No.~SH 81/3-1.
\end{acknowledgments}

\end{document}



\title{Supplemental Material for ``Dynamics of a Magnetic Needle Magnetometer: Sensitivity to Landau--Lifshitz--Gilbert Damping''}

\author{Y. B. Band$^{1,2}$, Y. Avishai$^{2,3,4}$, Alexander Shnirman$^{3,5,6}$}
\affiliation{
$^{1}$Department of Chemistry, Department of Physics, Department of Electro-Optics, and the Ilse Katz Center for Nano-Science, \\
Ben-Gurion University, Beer-Sheva 84105, Israel\\
$^{2}$New York University and the NYU-ECNU Institute of Physics at NYU Shanghai, 3663 Zhongshan Road North, Shanghai, 200062, China\\
$^{3}$Department of Physics, and the Ilse Katz Center for Nano-Science, \\
Ben-Gurion University, Beer-Sheva 84105, Israel\\
$^{4}$Yukawa Institute for Theoretical Physics, Kyoto, Japan\\
$^{5}$Institut f\"{u}r Theorie der Kondensierten Materie,
Karlsruhe Institute of Technology, D-76128 Karlsruhe, Germany\\
$^{6}$Institute of Nanotechnology, Karlsruhe Institute of Technology, D-76344
Eggenstein-Leopoldshafen, Germany
}


\maketitle

In this supplemental material we expand the discussion of the main text \cite{BAS} and address the following three issues. (a) The time $\tau_e$ at which the diffusion of the magnetic needle axis direction ${\bf n}$ and the magnetic needle orbital angular momentum ${\boldsymbol \ell}$ stops because equipartition is reached, i.e., we estimate the time required for the energy stored in stochastic orbital motion to become of order $k_B T$. (b) The uncertainty $\Delta B$ of the magnetic field for experiments in which the fast precession of ${\bf n}$ around ${\bf j}$ is averaged out in the measurement process and the uncertainty $\Delta B$ is determined by the diffusion of ${\bf j}$. (c) The dynamics of a magnetic needle in an inhomogeneous field, e.g., levitron dynamics of a rotating magnetic needle in a Ioffe-Pritchard trap \cite{Gov_Pritchard}, see Refs.~\cite{Berry, levitron_movie, Rusconi}.\\

(a): $\tau_e$ can be estimated by noting that the diffusion determined in \cite{BAS} stops once equipartition is reached.  The energy $\Delta E$ stored in stochastic orbital motion is given by
\begin{equation}
\Delta E \sim \hbar \omega_1 N \langle \delta {\boldsymbol \ell}^2 \rangle \ ,
\end{equation}
where where $N = S/\hbar$ (note that ${\bf \delta j} - {\bf \delta n} = \delta {\boldsymbol \ell}$).  By requiring $\Delta E \sim k_B T$ we can estimate that the diffusion given by Eqs.~(20-21) of \cite{BAS} stops when $\tau_e \sim \omega_0^2 /(\alpha \omega_1^3)$ (this result can also be obtained by expanding Eq.~(11) further in powers of $\lambda \equiv \omega_1/\omega_0$). For the parameters used in \cite{BAS} this is an extremely long time ($\tau_e \sim 10^{12}$ s $\sim  5$ years).  Hence, we conclude that the diffusion of Eqs.~(20-21) and the error estimates given for $\Delta B$ in Ref.~\cite{BAS} are relevant for all reasonable times.\\

(b): In \cite{BAS} we calculate $\Delta B$ assuming the experimental measurement follows the temporal dynamics of ${\bf n}$ and ${\bf j}$.  An alternative assumption is that the precession of ${\bf n}$ around ${\bf j}$ is averaged out by the measurement process and one measures the diffusion of ${\bf j}$. For the latter we obtain the leading term
\begin{equation}  \label{eq:djdiffusion1}
\langle \delta j_y^2(t) \rangle \approx  t\, \omega_B^2 \cos^2\theta \, S_a(\omega\sim \omega_1) \ ,
\end{equation}
where $S_a(\omega)$ is given in Eq.~(23) of \cite{BAS}.  At $\theta = \pi/2$ the leading contribution obtained in Eq.~(\ref{eq:djdiffusion1}) vanishes and the remaining sub-leading term is
\begin{equation}  \label{eq:djdiffusion2}
\langle \delta n_y^2(t) \rangle \approx  t\, \frac{2\omega_B^4}{\omega_1^2}\,  S_a(\omega \sim \omega_1) \ ,
\end{equation}
hence for $\theta \neq \pi/2$ we obtain
\begin{equation}
\Delta B = \frac{\Delta \omega_B}{\gamma}  \approx 
\frac{\hbar}{g\mu_B} \frac{\omega_B}{\omega_0}\, \cos\theta\,  \sqrt{\frac{2\alpha k_B T}{\hbar N}}\,\frac{1}{\sqrt{t}}\ ,
\end{equation}
whereas at $\theta = \pi/2$,
\begin{equation}
\Delta B = \frac{\Delta \omega_B}{\gamma}  \approx 
\frac{\hbar}{g\mu_B} \frac{\omega^2_B}{\omega_0\omega_1}\, \sqrt{\frac{4\alpha k_B T}{\hbar N}}\,\frac{1}{\sqrt{t}}\ .
\end{equation}
Taking $\omega_B = 1 \, {\rm s}^{-1}$ we obtain $\Delta B \approx \frac{\cos\theta \times 5 \times 10^{-23}}{\sqrt{t[s]}}$ Tesla for $\theta \neq \pi/2$, and $\Delta B \approx \frac{7 \times 10^{-25}}{\sqrt{t[s]}}$ Tesla for $\theta = \pi/2$.\\

(c): A rotating magnet can be levitated in an inhomogeneous magnetic field \cite{Berry, levitron_movie, Rusconi}.  This is possible despite Earnshaw's theorem \cite{Earnshaw} from which one can conclude that levitation of a {\em non-rotating} ferromagnet in a static magnetic field is not possible.  Two important factors regarding magnetic levitation are the forces on the magnet and its stability (ensuring that it does not spontaneously slide or flip into a configuration without lift).  The dynamics of a magnetic needle in an inhomogeneous magnetic field can be modelled using Eqs.~(6), (7) and (8) of \cite{BAS} augmented by the equations of motion for the center of mass (CM) degrees of freedom of the needle,
\begin{equation} \label{Eq:p}
{\dot {\bf p}} = {\boldsymbol \nabla} ({\boldsymbol \mu} \cdot {\bf B}({\bf r})) \ ,
\end{equation}
\begin{equation} \label{Eq:r}
{\dot {\bf r}} = {\bf p}/m \ ,
\end{equation}
where ${\bf r}$ and ${\bf p}$ are the needle CM position and momentum vectors.  Our numerical results show levitation of the magnetic needle when the initial rotational angular momentum vector of the needle is sufficiently large and points in the direction of magnetic field at the center of the trap. We shall see that the dynamical variables do not evolve to a fixed point or a simple cyclic orbit.  Moreover, a linear stability analysis yields a $15$$\times$$15$ Jacobian matrix with eigenvalues having a positive real part, so the system is {\em unstable}.  However, a stability analysis of the system using the adiabatic invariant $|{\boldsymbol \mu}| \, |{\bf B}|$ \cite{Berry} does yield a stable fixed point (contrary to the full numerical results which show a more complicated levitation dynamics).

\begin{figure} [ht]
\centering
\includegraphics[width=0.45\linewidth]{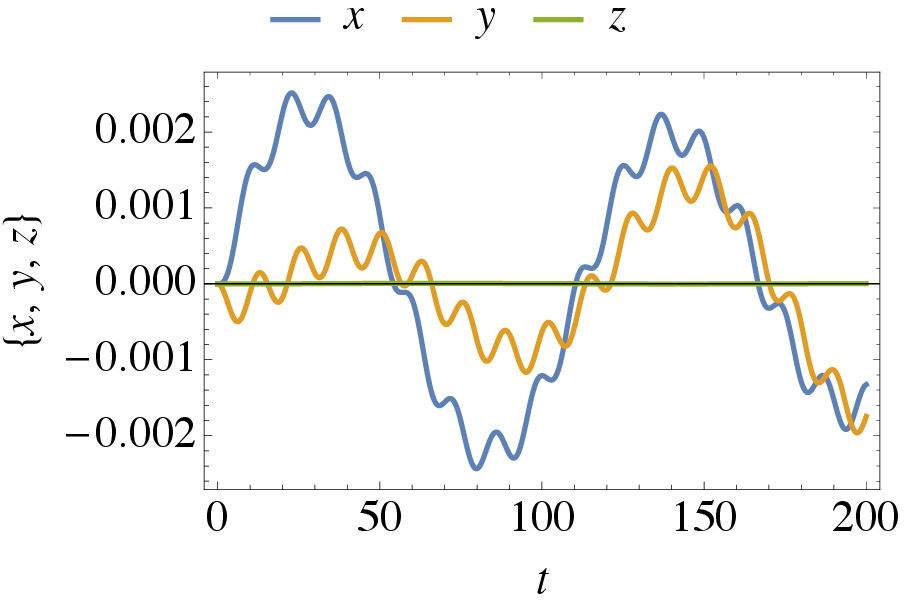}
\centering
\includegraphics[width=0.45\linewidth]{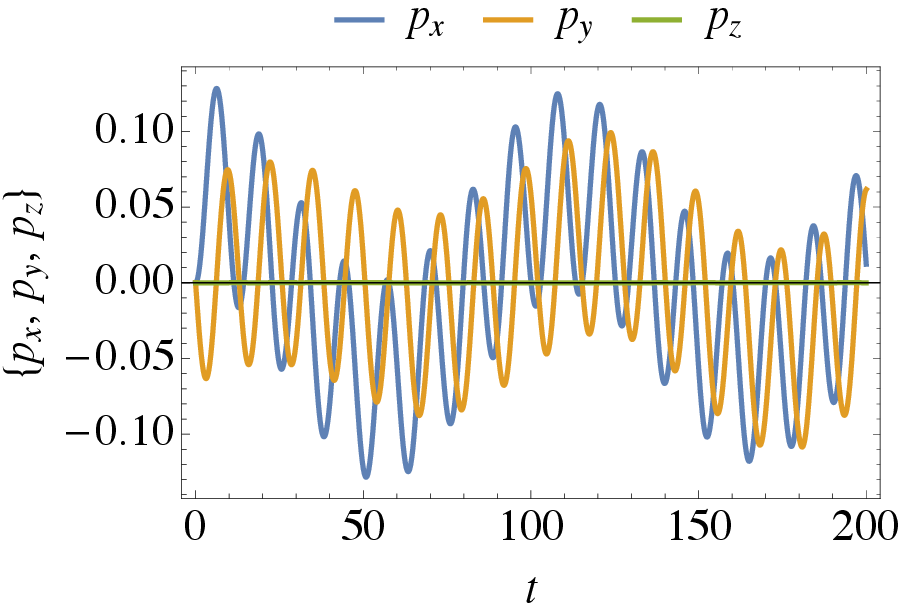}
\centering
\includegraphics[width=0.45\linewidth]{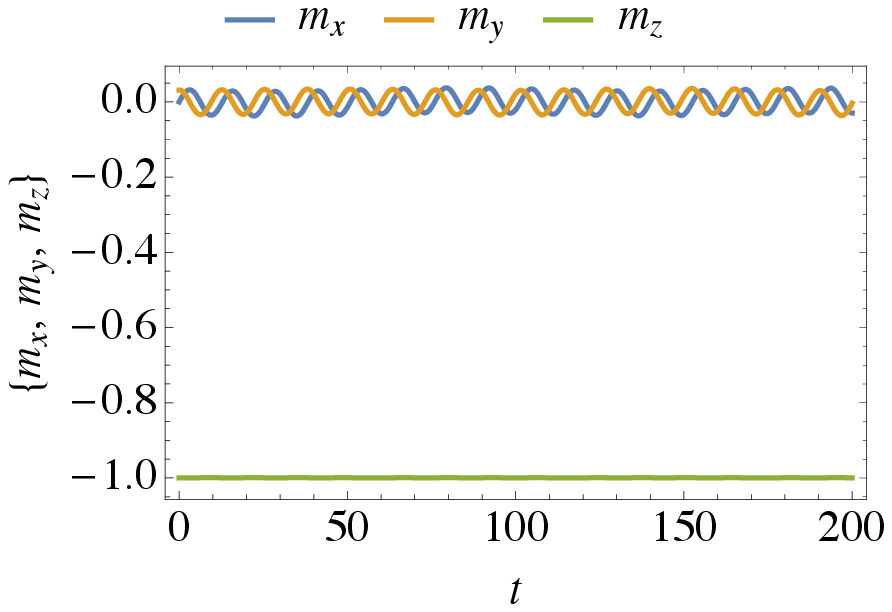}
\centering
\includegraphics[width=0.45\linewidth]{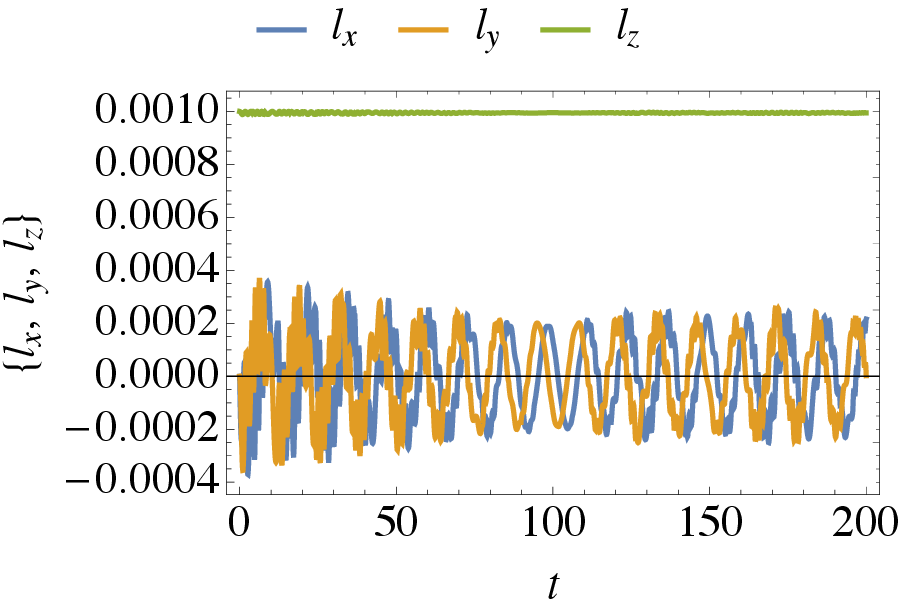}
\centering
\includegraphics[width=0.5\linewidth]{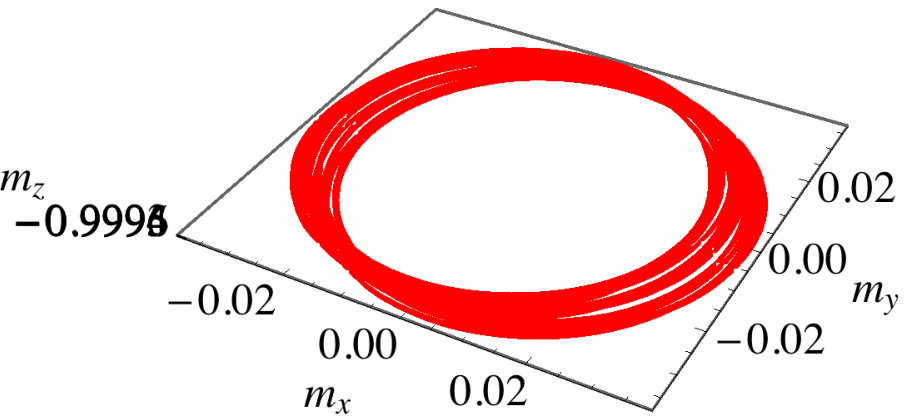}
\caption{(color online) Dynamics of a needle in a Ioffe-Pritchard magnetic field.  (a) ${\bf r}$ versus time, (b) ${\bf p}$ versus time, (c) ${\bf m}$ versus time (note that ${\bf n}(t)$ is indistinguishable from ${\bf m}(t)$ on the scale of the figure). (d) ${\boldsymbol \ell}$ versus time (note that $|{\boldsymbol \ell}(t)|$ is small compared to $S$ but rotational angular momentum ${\bf L}(t) = S \, {\boldsymbol \ell}(t)$ is large since $S = 10^{12}$).  (e) Parametric plot of the needle spin vector ${\bf m}(t)$ (nutation is very small for this case of small magnetic field).}
\label{Fig_1}
\end{figure}

Figure \ref{Fig_1} shows the dynamics of the system over time in the trap.  We use the same magnetic needle parameters used in Fig.~2 of \cite{BAS} and a Ioffe-Prichard magnetic field \cite{Gov_Pritchard}
\begin{equation} \label{Eq:B_IP}
{\bf B}({\bf r}) = {\bf e}_x \left( B' x - \frac{B''}{2} xz \right) + {\bf e}_y \left( B' y - \frac{B''}{2} zy \right) + {\bf e}_z \left( B_0 +  \frac{B''}{2} (z^2 - \frac{x^2 + y^2}{2}) \right) \ ,
\end{equation}
with field bias $B_0$, gradient $B'$, and curvature $B''$ parameters chosen so that the Zeeman energy and its variation over the trajectory of the needle in the trap are substantial (as is clear from the results shown in the figure).  
We start the dynamics with initial conditions: ${\bf r}(0) = (0,0,0)$, ${\bf p}(0) = (0,0,0)$, ${\bf m}(0) = (0,0.001^{1/2},-(1-0.001)^{1/2})$ (almost along the $-z$ direction), ${\bf n}(0) = {\bf m}(0)$, ${\boldsymbol \ell}(0) = (0,0,0.001)$ [this is large orbital angular momentum since ${\boldsymbol \ell}$ is the orbital angular momentum divided by $S$].  Figure \ref{Fig_1}(a) shows the needle CM position ${\bf r}(t)$ versus time.  Fast and slow oscillations are seen in the $x$ and $y$ motion, whereas $z(t)$ remains very close to zero. Figure~\ref{Fig_1}(b) shows oscillations of the CM momentum ${\bf p}(t)$ with time.  $p_x(t)$ and $p_y(t)$ oscillate with time, and $p_z(t)$ remains zero.  Figure~\ref{Fig_1}(c) plots the spin ${\bf m}(t)$ versus time.  Initially, ${\bf m}(0)$ points almost in the $-z$ direction, and the tip of the needle ${\bf n}(t) = {\bf m}(t)$ carries out nearly circular motion in the $n_x$-$n_y$ plane.  Figure~\ref{Fig_1}(d) plots the orbital angular momentum ${\boldsymbol \ell}(t)$. The components $\ell_x(t)$ and $\ell_y(t)$ undergo a complicated oscillatory motion in the $\ell_x(t)$-$\ell_y(t)$ plane but $\ell_z(t) \approx \ell_z(0)$.  Figure~\ref{Fig_1}(e) is a parametric plot of ${\bf m}(t)$; the motion consists of almost concentric rings that are slightly displaced one from the other.  The full dynamics show levitation but they do not converge to a fixed point or a limit cycle.

Quite generally, for a system of dynamical equations, $\dot y_i(t) = f_i(y_1, \ldots, y_n)$, $i = 1, \ldots n$, a linear stability analysis requires calculating the eigenvalues of the Jacobian matrix evaluated at the equilibrium point ${\bf y}^*$ where ${\bf f}({\bf y}^*)= {\bf 0}$, $J_{ij} = \left(\frac{\partial f_i}{\partial y_j}\right)_{{\bf y}^*}$ \cite{Merkin}.  The system is unstable against fluctuations if any of the eigenvalues of $J_{ij}$ have a positive real part. Equations~(6), (7) and (8) of \cite{BAS} together with Eqs.~(\ref{Eq:p}) and (\ref{Eq:r}) above have a Jacobian matrix with eigenvalues whose real part are positive, so the linear stability test fails.  However, if the Zeeman force $-{\boldsymbol \nabla} H_Z$ in Eq.~(\ref{Eq:p}) is replaced by the gradient of the adiabatic invariant, ${\boldsymbol \mu} \cdot {\boldsymbol \nabla} |{\bf B}({\bf r})|$, none of the eigenvalues of the Jacobian matrix have a positive real part and the system is linearly stable, i.e., the stability analysis using the adiabatic-invariant predicts stability.  Note that substituting the adiabatic invariant for the Zeeman energy in the full equations of motion yields ${\bf r}(t)$ and ${\bf p}(t)$ vectors that are constant with time and ${\bf n}(t)$, ${\bf m}(t)$ and ${\boldsymbol \ell}(t)$ are similar to the results obtained with the full equations of motion (but the parametric plot of ${\bf m}(t)$ is a perfectly circular orbit).  Thus, adiabatic--invariant stability analysis of a rotating magnetic needle in a magnetic trap confirms stability of its levitation as obtained in the numerical solution of the dynamical equations.

